\documentclass[onecolumn,showpacs,12pt]{revtex4}
\usepackage{graphics,epsfig,graphicx}
\usepackage{color}
\usepackage{makeidx}
\usepackage{latexsym}

\begin{document}

%======================================%
%<<<<<<<<<<<< TITLE PAGE >>>>>>>>>>>>>>%
%======================================%

\title{Role of Fermion Exchanges in Statistical Signatures of\\ Composite Bosons}

\author{M. Combescot$^{(1)}$, F. Dubin$^{(2)}$ and M. A. Dupertuis$^{(3)}$}
\affiliation{$^{(1)}$ INSP-Institut des NanoSciences de Paris, Universit\'{e} Pierre et Marie Curie - CNRS, 140 rue de Lourmel, 75015 Paris\\
$^{(2)}$ ICFO-Institut de Ci\`{e}ncies Fot\`{o}niques, Mediterranean Technology Park, E-08860 Castelldefels, Spain\\
$^{(3)}$ Laboratoire d'Opto\'{e}lectronique Quantique et Laboratoire de Physique des Nanostructures,
Ecole Polytechnique F\'{e}d\'{e}rale de Lausanne EPFL, Station 3, CH-1015 Lausanne, Switzerland}
\date{\today}

\begin{abstract}
We study statistical signatures of composite bosons made of two fermions using a new many-body approach. Extending number-states to composite bosons, two-particle correlations as well as the dispersion of the probability distribution are analyzed. We show that the particle composite nature reduces the anti-bunching effect predicted for elementary bosons. Furthermore, the probability distribution exhibits a dispersion which is greater for composite bosons than for elementary bosons. This dispersion corresponds to the one of sub-Poissonian processes, as for a quantum state, but, unlike its elementary boson counterpart, it is not minimum. In general, our work shows that it is necessary to take into account the Pauli exclusion principle which takes place between fermionic components of composite bosons - along the line here used - to possibly extract statistical properties in a precise way. 
\end{abstract}

\maketitle

% Introduction part

Quantum particles obey statistical laws that are highlighted by high-order correlations. This has first been shown for light fields, by the discovery of photon bunching in two-photon correlations from an incoherent source \cite{HBT}. Thereafter, second-order correlations have become increasingly important, particularly to distinguish classical from quantum states of light \cite{MWolf,Scully}. Indeed, classical states are characterized by bunched two-photon correlations while quantum states yield anti-bunching. For elementary bosons, two-particle correlations are directly connected to the variance of the probability distribution which is usually expressed through the so-called Mandel parameter $Q$ \cite{Mandel1979}. The latter compares the dispersion of the distribution to the one of Poissonian processes. Hence, quantum states  correspond to negative values of $Q$, i.e., they follow sub-Poissonian statistics, while classical states lead to positive values of $Q$ since these can solely follow Poissonian and super-Poissonian statistics.

Recent advances with ultra-cold atoms have shown that analysis of second-order correlations is also well suited to probe matter states (see \cite{Bloch2008} and references therein). Most notably, anti-bunching and bunching effects have been observed while measuring two-particle correlations of degenerate Fermi and Bose gases \cite{Fohling2005,Rom2006,Jeltes2007}. In quantum mechanical terms, these effects are due to interfering amplitudes for the two paths that particles can take to reach the detectors. Intimately, bunching and anti-bunching reflect the statistics obeyed by the quantum particles: Bose-Einstein statistics imposes addition of the amplitudes, i.e., constructive interference, while Fermi-Dirac statistics yields destructive interference. In the latter case, as for elementary bosons, an anti-bunching effect signals that the matter state has no classical analog. Note that fermion anti-bunching was also observed for electrons \cite{Oliver1999,Henry1999,Kiesel2002} and neutrons \cite{Iannuzzi2006}. 

One encounters a more complex situation while studying composite bosons made of two fermions. In an ensemble of such quantum particles, the Pauli exclusion principle induces fermion exchanges between composite bosons which therefore do not exactly follow Bose-Einstein statistics. Semiconductor excitons constitute a good example of such composite bosons. These are made of electron-hole pairs and fermion exchanges between the electrons and holes of these excitons are highlighted by optical nonlinearities \cite{SC}. Interacting Fermi gases, made of  \textit{e.g.} $^6$Li and $^{40}$K, constitute other systems where composite bosons can be studied. In ultra-cold samples \cite{11}, these offer an interesting route towards precise investigation of the Pauli exclusion principle. Indeed, exchanges between fermionic components can be varied, for instance when imbalanced spin populations
interact \cite{12}.

Recently, a significant breakthrough was made in the
theoretical description of composite bosons made of two fermions. A novel formalism, free from any mapping to an ideal boson subspace, has been constructed, and a visualization of the physical processes taking place
between composite bosons (cobosons) has been proposed through the so-called ``Shiva'' diagrams (see \cite{report} for a general review). This coboson many-body theory relies on two sets of $2\times 2$ scatterings: the
``interaction scatterings'' and the ``Pauli scatterings''. The first ones correspond to interactions
between the fermionic components of two cobosons in the absence of fermion exchanges while Pauli scatterings correspond to fermion exchanges between two
cobosons, without any fermion interaction. 

In this work, we use this new many-body approach to study the influence of fermion exchanges in statistical signatures of composite bosons. To highlight corresponding effects, we extend number-states to composite bosons by considering a many-body state constructed from $N$ identical coboson creation operators. Hence, we evaluate the number of coincidences in two-particle correlations, $g^{(2)}_N$, as well as the variance of the field which is calculated through the Mandel parameter $Q_N$. 

For number-states made of elementary bosons, $|\bar{\psi}_N\rangle=\bar{B}_0^{\dag N}|v\rangle$, where $\bar{B}_0^{\dag}$ is the field-particle creation operator and $|v\rangle$ the vacuum state, it is known that $\bar{g}^{(2)}_N$ and $\bar{Q}_N$ read as (1-1/$N$) and (-1) respectively: Number-states exhibit an anti-bunched second-order correlation function and yield the greatest negative value for the Mandel parameter. This implies that measurements of such states are made with the greatest sensitivity, i.e., that noise is reduced at the smallest level.  For composite bosons in a many-body state $|\psi_N\rangle=B_0^{\dag N}|v\rangle$, we find that fermion exchanges modify the number of coincidences which now reads $g^{(2)}_N$$\simeq 1+[-1+O(\eta)]/N$ in the large $N$ limit. $O(\eta)$ is a positive correction whose dominant term increases linearly with the dimensionless parameter $\eta=N(a_B/L)^D$ associated to the composite boson density, $a_B$ being the coboson extension, $L$ the sample size and $D$ the space dimension. Consequently, two-particle correlations of composite bosons present a reduced anti-bunching effect when compared to elementary bosons. Furthermore, the amplitude of the $Q$-parameter for the state  $|\psi_N\rangle$ reads as $Q_N\simeq-1+O(\eta^2)$. Hence, the dispersion of the probability distribution is increased by fermion exchanges. In general, our analysis shows that it is necessary to take into account the Pauli exclusion principle between fermionic components of composite bosons in order to possibly extract statistical properties of these quantum particles in a precise way.

\section{Physical Understanding}

Statistical properties of quantum fields are often studied through the second-order correlation function. The latter evaluates fluctuations in the number of field-particles for a given many-body state. Precisely, it measures the normalized probability to detect a particle conditioned upon detection of a previous particle. The normalized number of two-particle coincidences in a state 0 is defined as
\begin{equation}
g^{(2)}_N=\frac{\langle B_0^{\dag 2}B_0^2 \rangle_N}{\langle B_0^{\dag}B_0\rangle_N^2}\label{eq1},
\end{equation}
where $\langle A \rangle_N=\langle\psi_N|A|\psi_N\rangle$/$\langle\psi_N|\psi_N\rangle$ is the mean value of the operator $A$ in state $|\psi_N\rangle$, the field-particle creation operator being  $B_0^\dag$. 

For elementary bosons, comparison between the normalized number of coincidences and 1 allows us to deduce whether fluctuations of the number of particles follow classical or quantum statistics. Another way to study the variance of the number of particles and to characterize a quantum field is via the so-called Mandel parameter which precisely reads 
\begin{equation}
Q_N=\frac{\langle\hat{n}^2\rangle_N-\langle \hat{n}\rangle_N^2}{\langle \hat{n}\rangle_N}-1\label{eq2},
\end{equation}
where $\hat{n}=B^\dag_0B_0$ is the number-operator for the state $0$. 

The parameter $Q_N$ compares the fluctuations of the field number-operator to that of a Poissonian source \cite{Mandel1979}. A negative $Q_N$ signals that the field statistics is sub-Poissonian and hence corresponds to the one of a quantum state. Most striking examples are obtained for the number-states $|\bar{\psi}_{N}\rangle=\bar{B}_0^{\dag N}|v\rangle$ of particles with bosonic statistics, i.e. $[\bar{B}_m,\bar{B}^\dag_i]= \delta_{mi}$. For these, we find $\bar{g}^{(2)}_N=(1-1/N)$ and $\bar{Q}_N=-1$ for any particle $0$. Note that $(-1)$ is the greatest possible negative value allowed for the Mandel parameter since the ratio in Eq.(2) is always positive: its denominator is the norm of ${B_0}|{\psi}_{N}\rangle$ while its numerator is the norm of $P_{\perp}\hat{n}|{\psi}_{N}\rangle$ where $P_{\perp}=1-|{\psi}_{N}\rangle\langle{\psi}_N|/\langle {\psi}_{N}|{\psi}_{N}\rangle$ is the projector over the subspace perpendicular to $|{\psi}_{N}\rangle$, this operator being such that $P_{\perp}=P_{\perp}^2$. 

To demonstrate that fermion exchanges play an important role in statistical signatures of composite bosons, we extend boson number-states to cobosons and consider the unnormalized ket  $|\psi_{N}\rangle=B_0^{\dag N}|v\rangle$ to evaluate $g^{(2)}_N$ and $Q_N$. In order to understand how fermionic components affect these two quantities in a simple way, let us start with $N$=2 since most physical effects induced by the Pauli exclusion principle already appear with just two composite particles.

\subsection{Two composite bosons} 

Since $B_0^2|\psi_2\rangle$ is a zero-pair state, $\langle \psi_2|B_0^{\dag 2}B_0^2|\psi_2 \rangle$
is equal to $\langle\psi_2|B_0^{\dag 2}|v\rangle\langle v|B_0^{2}|\psi_2\rangle$; so that by noting that 
$\langle v|B_0^{2}|\psi_2\rangle$ is nothing but $\langle\psi_2|\psi_2\rangle$, the normalized two-particle coincidences reduce to
\begin{equation}
g^{(2)}_2= \frac{\langle\psi_2|\psi_2 \rangle^3}{\langle\psi_2|\hat{n}|\psi_2 \rangle^2}.\label{eq3}
\end{equation}

To better see differences induced by the particle composite nature, let us briefly reconsider elementary bosons, i.e., bosons such that $[\bar{B}_m,\bar{B}^\dag_i]= \delta_{mi}$. We then have $\bar{B}_0|\bar{\psi}_2\rangle$=$2|\bar{\psi}_1\rangle$ so that $\bar{B}_0^2|\bar{\psi}_2\rangle$= 2$|v\rangle$ and $\hat{n}|\bar{\psi}_2\rangle= 2|\bar{\psi}_2\rangle$. Consequently $\langle\bar{\psi}_2|\bar{\psi}_2\rangle=2$ while $\langle\bar{\psi}_2|\hat{n}|\bar{\psi}_2\rangle=4$ and $\langle\bar{\psi}_2|\hat{n}^2|\bar{\psi}_2\rangle=8$. This leads to the expected results, namely $\bar{Q}_2= -1$ and $\bar{g}^{(2)}_2= 1/2$ which is nothing but $(1-1/N)$ taken for $N=2$.

We now turn to composite bosons. These have creation operators which do not exactly follow bosonic commutation rules \cite{report}, but instead
\begin{equation}
[B_m,B^\dag_i]= \delta_{mi}-D_{mi},\label{eq4}
\end{equation}
where the so-called ``deviation-from-boson'' operator, $D_{mi}$, is such that $D_{mi}|v\rangle=0$ while 
\begin{equation}
[D_{mi},B^\dag_j] = \sum_n \left(\lambda(^{n \hspace{.2cm} i}_{m\hspace{.15cm} j})
+i \leftrightarrow j \right)B^\dag_n.\label{eq5}
\end{equation}
The parameters $\lambda(^{n \hspace{.2cm} i}_{m\hspace{.15cm} j})$ are the composite bosons "Pauli scatterings"; they describe fermion exchanges between ($i,j$) in the absence of fermion interaction (see Fig. 1.(a)).

\begin{figure}
\includegraphics[width=\textwidth]{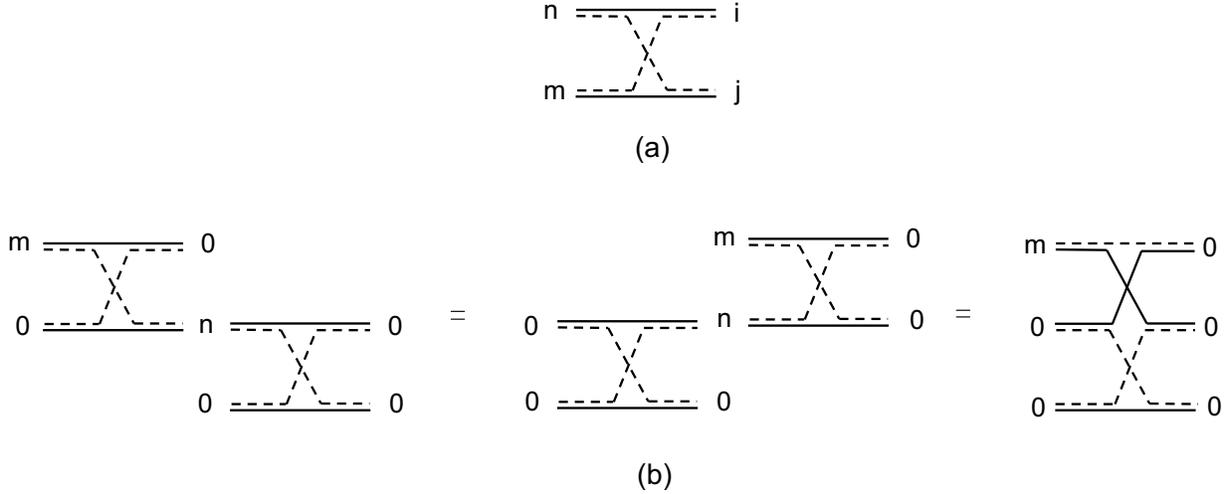}
\caption{Shiva diagrams representing fermion exchanges between composite bosons. Fermionic components are represented by solid and dashed lines. (a): Fermion exchange between two "in" cobosons in states $(i,j)$ leading to the formation of two "out" cobosons in states $(n,m)$. This diagram represents the Pauli scattering $\lambda(^{n \hspace{.21cm} i}_{m\hspace{.15cm} j})$. (b): Fermion exchanges between three composite bosons. Two possible ways to decompose fermion exchanges between three "in" cobosons $0$ and three "out" cobosons $(m,0,0)$ are represented, as appearing in Eqs (7) and (38). For $m=0$, this corresponds to the three-body scattering $\lambda_3 = \sum_n \lambda(^{0 \hspace{.15cm} 0}_{0\hspace{.15cm} n})\lambda(^{n \hspace{.15cm} 0}_{0\hspace{.15cm} 0})$ appearing in Eq.(8). }
\end{figure}

The previous commutators give $B_0|\psi_2\rangle= 2|\psi_1\rangle$-$2L_2^\dag|v\rangle$ and $B_0L_2^\dag|\psi_1\rangle = \lambda_2|\psi_1\rangle+(L_2^\dag-2L^\dag_3)|v\rangle$ where we have set  $\lambda_2 = \lambda(^{0 \hspace{.15cm} 0}_{0\hspace{.15cm} 0})$ while
\begin{eqnarray}
L_2^\dag = \frac{1}{2}[D_{0 0},B^\dag_0] = \sum_n \lambda(^{n \hspace{.2cm} 0}_{0\hspace{.15cm} 0})B^\dag_n\label{eq6}\\
L_3^\dag = \frac{1}{2}[D_{0 0},L_2^\dag] = \sum_{n,p} \lambda(^{p \hspace{.2cm} 0}_{0\hspace{.15cm} n})\lambda(^{n \hspace{.2cm} 0}_{0\hspace{.15cm} 0})B^\dag_p . \label{eq7}
\end{eqnarray}
This allows one to find $B_0^2|\psi_2\rangle = 2(1-\lambda_2)|v\rangle$ such that $\langle\psi_2|\psi_2\rangle = 2(1-\lambda_2)$. This also leads to $\hat{n}|\psi_2\rangle = 2|\psi_2\rangle - 2L_2^\dag|\psi_1\rangle$. In order to calculate  $\langle \psi_2|\hat{n}|\psi_2 \rangle$ in an easy way, it is convenient to first note that $\langle\psi_2|L^\dag_2|\psi_1\rangle = \langle\psi_2|B_0^\dag L_2^\dag|v \rangle$. From $\langle \psi_2|B_0^\dag$ as given above, we then find that $\langle \psi_2|L_2^\dag|\psi_1 \rangle = 2(\lambda_2-\lambda_3)$ where $\lambda_3 = \sum_n \lambda(^{0 \hspace{.15cm} 0}_{0\hspace{.15cm} n})\lambda(^{n \hspace{.15cm} 0}_{0\hspace{.15cm} 0})$ corresponds to the Shiva diagram for fermion exchanges between three composite bosons 0, shown in Figure 1.(b). All this leads to 
\begin{equation}
\langle\psi_2|\hat{n}|\psi_2 \rangle = 4(1-2\lambda_2+\lambda_3).\label{eq8}
\end{equation}
 The most convenient way to obtain 
  $\langle\psi_2|\hat{n}^2|\psi_2\rangle$ is to first calculate $B_0\hat{n}|\psi_2\rangle$ knowing $B_0L_2^\dag|\psi_1\rangle$ given above. By noting that $\langle \psi_2|L_3^\dag|\psi_1\rangle = 2(\lambda_3-\lambda_4)$, where $\lambda_4$ corresponds to the Shiva diagram for fermion exchanges between four composite bosons $0$, we end with 
\begin{equation}
\langle \psi_2|\hat{n}^2|\psi_2 \rangle = 4 (2-6\lambda_2+\lambda_2^2+5\lambda_3-2\lambda_4).\label{eq9}
\end{equation}
Before going further, we can check that the above results for cobosons reduce to the ones for elementary bosons when all fermion exchanges, i.e. all $\lambda_n$'s, are dropped. 

For cobosons $0$ with center of mass momentum \textbf{Q$_0$} and relative motion index $\nu_0$, it is possible to show \cite{report} that  
$\lambda_n=\sum_{\textbf{k}} |\langle \textbf{k}|\nu_0 \rangle|^{2n}$. This makes $\lambda_n$ a positive constant of the order of $(a_B/L)^{(n-1)D}$ where $a_B$ is the spatial extension of the relative motion state $|\nu_0\rangle$ and $D$ the space dimension. For 3D excitons or Hydrogen atoms, for which $|
\langle \textbf{k}|\nu_0 \rangle|^2$=$64\pi(a_B/L)^3/(1+k^2a_B^2)^2$, we find
\begin{equation}
\lambda_n = 16 \frac{(8n-5)!!}{(8n-2)!!}\left( \frac{64\pi a_B^3}{L^3} \right)^{n-1}.\label{eq12}
\end{equation}
This leads to $\lambda_2$=$(33\pi/2)(a_B/L)^3$ and $\lambda_3$=$(4199\pi^2/8)(a_B/L)^6$ which makes $\lambda_3$$\ll$$\lambda_2$ and ($\lambda_3$-$\lambda_2^2$) appearing below, in the Mandel parameter, a positive constant equal to $(2021\pi^2/8)(a_B/L)^6$. 

 By collecting all the above results, we end with a normalized number for two-particle coincidences of composite bosons which reads as
\begin{equation}
g^{(2)}_2= \frac{1}{2}\frac{(1-\lambda_2)^3}{(1-2\lambda_2+\lambda_3)^2}\simeq \frac{1}{2}(1+\lambda_2),\label{eq10}
\end{equation}
while the Mandel parameter is given by 
\begin{equation}
 Q_2\approx -1+(\lambda_3-\lambda_2^2).\label{eq11}
\end{equation}

This shows, for $N=2$, that the number of two-particle coincidences is slightly higher for composite bosons than for elementary ones. We also find that the Mandel parameter is increased. As seen from Eqs.(\ref{eq10},\ref{eq11}), the corresponding variations are solely controlled by the Pauli exclusion principle, through the exchange scatterings $\lambda_n$ between $n$ cobosons $0$. Unlike elementary bosons for which the second-order correlation function and the Mandel parameter depend on $N$ only, their counterparts for composite bosons also depend on the state $0$ at hand.

\subsection{Guess for N composite bosons}

If we now turn to $N$ elementary bosons, number-states are such that $\hat{n}|\bar{\psi}_N\rangle= N|\bar{\psi}_N\rangle$ such that
$\bar{g}^{(2)}_N=1-1/N$ and $\bar{Q}_N=-1$ for all $N$, whatever the state $0$ is. 

For composite bosons, a "rule of the thumb" gives intuitively the leading term in the small density limit, through the replacement of $(a_B/L)^D$ by $(N-1)(a_B/L)^D$ in the result for $N= 2$. Since $\lambda_3$ and $\lambda_2^2$ are both of the order of (a$_B$/L)$^D$, we are led to guess the Mandel parameter as
\begin{equation}
Q_N\simeq -1+(N-1)^2(\lambda_3-\lambda_2^2) \simeq -1+O(\eta^2),\label{eq13}
\end{equation}
where $\eta$ is the previously defined dimensionless parameter associated to density. The situation for $g^{(2)}_N$ is more ambiguous. From Eq.(\ref{eq10}), we might guess 
\begin{equation}
g^{(2)}_N \approx \left(1-\frac{1}{N}\right)\left(1+(N-1)\lambda_2\right)\approx \left(1-\frac{1}{N}\right)\left(1+O(\eta)\right ),\label{eq14a}
\end{equation}
This would lead to a crossover from anti-bunching to bunching when $1/N$ crosses $\eta$, i.e., when the system size increases - which is very unlikely physically. However, we might as well guess
\begin{equation}
g^{(2)}_N\simeq1+\frac{-1+(N-1)\lambda_2}{N}\simeq 1+\frac{-1+O(\eta)}{N}\label{eq14b},
\end{equation}
which maintains the anti-bunching effect for all sample sizes since the $\lambda_n$ expansions performed in these calculations only hold for $\eta\ll$ 1. The purpose of the next section is to demonstrate that Equations (\ref{eq13}) and (\ref{eq14b}) are indeed the correct expressions of $Q_N$ and $g^{(2)}_N$. To do it, we must go deeper into the composite boson many body theory.

\section{ Second-order correlations for N composite bosons}

\subsection{Key results from the coboson many-body theory}

Many-body effects between $N$ composite bosons linked to fermion exchanges are obtained through the two following commutators \cite{report}

\begin{equation}
[B_m,B_i^{\dag N}]= NB_i^{\dag N-1}(\delta_{mi}-D_{mi})-N(N-1)B_i^{\dag N-2}
\sum_n \lambda(^{n \hspace{.15cm}i}_{m\hspace{.1cm} i})B_n^\dag\label{eq15},
\end{equation}
\begin{equation}
[D_{mi},B_j^{\dag N}]= NB_j^{\dag N-1}\sum_n \left( \lambda(^{n \hspace{.15cm}j}_{m\hspace{.1cm} i})+\lambda(^{n \hspace{.15cm}i}_{m\hspace{.1cm} j})\right)B_n^\dag\label{eq16}.
\end{equation}
Note that these two commutators reduce to Eqs.(\ref{eq4},\ref{eq5}) when $N=1$. 

Using them, it is possible to show that the normalization factor for $N$ cobosons in the same $0$ state, differs from its elementary bosons value ($N!$), due to fermion exchanges taking place between them. This led us to write 
\begin{equation}
\langle v|B_0^NB_0^{\dag N}|v\rangle=N!F_N,\label{eq17}
\end{equation}
where the $F_N$'s, which enter all calculations involving a large number of identical cobosons, follow the recursion relation
\begin{eqnarray}
F_N= F_{N-1}-(N-1)\lambda_2F_{N-2}+(N-1)(N-2)\lambda_3F_{N-3}-...\nonumber\\
\hspace{-1cm}=\sum_n (-1)^{n-1}\frac{(N-1)!}{(N-n)!}\lambda_nF_{N-n}.\nonumber\\
\label{eq18}
\end{eqnarray}
The $\lambda_n$'s are the previously defined scatterings for fermion exchanges between $N$ cobosons $0$, as shown in Fig. 1 for $N$=2 or 3.

Equation (\ref{eq15}) readily shows that 
\begin{equation}
%D_{00}|\psi_N\rangle = 2N L_2^\dag |\psi_{N-1}\rangle,\label{eq19}\\
B_0 |\psi_N\rangle = N |\psi_{N-1}\rangle -N(N-1)L_2^\dag |\psi_{N-2}\rangle, \label{eq20}\\
%L_2|\psi_N\rangle = N\lambda_2|\psi_{N-1}\rangle - N(N-1)L_3^\dag|\psi_{N-2}.\label{eq21}
\end{equation}
with $L_2^\dag$ given in Eq.(6). The above equation used for $B_0|\psi_N\rangle$ and then for $L_2^\dag|\psi_{N-1}\rangle$, allows us to show that
\begin{equation}
B_0^\dag B_0 |\psi_N\rangle = |\psi_N\rangle +\frac{N-1}{N+1}B_0|\psi_{N+1}\rangle.\label{eq22}
\end{equation}
This last equation will turn very useful in calculating matrix elements involving the number-operator. 

\subsection{Calculation of $Q_N$}

Since $\langle \psi_N|B_0 = \langle \psi_{N+1}|$, Eqs.(\ref{eq17}) and (\ref{eq22}) give the mean value of the particle 0 number-operator as
\begin{equation}
\langle \hat{n} \rangle_N = 1+(N-1)\frac{F_{N+1}}{F_N}=N+(N-1)\Delta_N^{(1)},\label{eq23}
\end{equation}
where we have set $\Delta_N^{(n)}= (F_{N+n}-F_{N+n-1})/F_N$.

For $N=2$ cobosons, Eq.(\ref{eq18}) gives $F_2 = 1-\lambda_2$ and $F_3 = 1-3\lambda_2+2\lambda_3$, so that the above equation agrees with Eq. (\ref{eq8}). For $N$ elementary bosons, $F_N$ reduces to 1 and consequently all the $\Delta_N^{(n)}$ differences reduce to zero. Hence, the number operator mean value reduces to $N$, as expected. By contrast, $\Delta_N^{(n)}$ for composite bosons are negative scalars since $F_N$ is a decreasing function of $N$, as seen from Eq.(\ref{eq18}). This makes the number operator mean value for cobosons smaller than its elementary boson value $N$, due to a "moth-eaten" effect similar to the one we have already found in other problems dealing with composite bosons: when a coboson 0 is added to $N$ other cobosons 0, the additional coboson feels the other $N$'s through the Pauli exclusion principle. Therefore $N$ elementary fermion pair states are blocked and thus missed in its linear combination, as if these were "eaten" by $N$ "moths". This picture allows us to physically understand all decreases found for cobosons, when compared to their counterpart elementary boson values. 

By repeatedly using Eq.(\ref{eq22}) for $\langle \psi_N|\hat{n}^2|\psi_N\rangle$, split as $(\langle \psi_N|\hat{n})(\hat{n}|\psi_N\rangle)$, we can show that
\begin{equation}
\langle \hat{n}^2\rangle_N = N^2+(N^2-1)\Delta_N^{(1)}+\frac{N(N-1)^2}{N+1}\Delta_N^{(2)}\label{eq24}.
\end{equation}
Again, $\langle \hat{n}^2 \rangle_N$ for cobosons is smaller than its $N^2$ value for elementary bosons, due to the same moth-eaten effect, since all the $\Delta_N^{(n)}$ are negative. Using $F_4 = 1-6\lambda_2+8\lambda_3+3\lambda_2^2-6\lambda_4$, as deduced from Eq. (\ref{eq18}), it is possible to check that the above equation agrees with Eq.(\ref{eq9}) when $N= 2$.

The two above equations allow us to write the mean quadratic deviation in a compact form in terms of two $F_N$ differences only, namely
\begin{equation}
\langle \hat{n}^2\rangle_N - \langle \hat{n}\rangle_N^2 = (N-1)^2[-\Delta_N^{(1)}(1+\Delta_N^{(1)})+\frac{N}{N+1}\Delta_N^{(2)}].\label{eq25}
\end{equation}
Since, in the small density limit, $\Delta_N^{(1)}\approx -N\lambda_2+N(N-1)(\lambda_3-\lambda_2^2)$ and $\Delta_N^{(2)}\approx -(N+1)\lambda_2+(N+1)N\lambda_3$, as deduced from Eq.(\ref{eq18}), we find that the mean quadratic deviation has no term in $(a/L)^D$, its small density dominant term being $N(N-1)^2(\lambda_3-\lambda_2^2)$. Using Eqs. (\ref{eq23}), it is then easy to show that the small density value of $Q_N$ given in Eq. (\ref{eq13}), as guessed from the rule of the thumb, is fully correct.

\subsection{Calculation of $g^{(2)}_N$}

The calculation of the normalized number of coincidences is a little more demanding. To evaluate the numerator of $g^{(2)}_N$, we first rewrite $B_0^\dag B_0$ using the commutator (\ref{eq4}). This yields

\begin{equation}
 \langle B_0^\dag(B_0^\dag B_0)B_0^\dag \rangle_N = \langle \hat{n}^2-\hat{n} \rangle_N+\langle B_0^\dag D_{00}B_0 \rangle_N.\label{eq26}
\end{equation}
The first term readily follows from Eqs.(\ref{eq23},\ref{eq24}). To express the second term in a compact form is more difficult. Non-trivial manipulations are indeed necessary, otherwise we end with $F_N$ expansions which are far from obvious to sum. These manipulations are reported in the Appendix. They lead to
\begin{eqnarray}
\langle B_0^\dag D_{00}B_0 \rangle_N = -2[\lambda_2+\Delta_N^{(1)}+\frac{(N-1)^2}{N+1}\Delta_N^{(2)}]\nonumber\\
= -2N(N-1)\left(\frac{\Delta_N^{(2)}}{N+1} -R \right)\label{eq27},
\end{eqnarray}
where $R$, defined as
\begin{equation}
R = \frac{1}{N}\frac{\Delta_N^{(2)}}{N+1} -\frac{1}{N(N-1)}(\lambda_2+\Delta_N^{(1)}),\label{eq30}
\end{equation}
 tends to $\lambda_2^{2}$ in the large sample limit (see Appendix).

Using Eqs.(22,23,25,26), it is then easy to show that 
\begin{equation}
\langle B_0^{\dag 2}B_0^2 \rangle_N = N(N-1)\left[ 1+\Delta_N^{(1)}+\frac{N-3}{N+1}\Delta_N^{(2)}+2R \right]\label{eq29}.
\end{equation}
If we now remember that, in the large sample limit, $\Delta_N^{(1)}$ tends to $-N\lambda_2$, $\Delta_N^{(2)}$ to $-(N+1)\lambda_2$ and  $R$ to $\lambda_2^2$, we find that the bracket in the above equation tends to $[1-(2N-3)\lambda_2]$. We then deduce using Eq.(22) that $g_N^{(2)}$ is given by \begin{equation}
g_N^{(2)}\approx (1-\frac{1}{N})(1+\lambda_2)\approx 1+\frac{-1+N\lambda_2}{N},\label{eq32}
\end{equation}
in agreement with the result guessed in Eq.(\ref{eq14b}), since $(N-1)\approx N$ in the large $N$ limit.

\section{Conclusions}

By extending number-states to composite bosons, we have shown that the Pauli exclusion principle modifies statistical signatures of composite bosons many-body quantum states. This is underlined through the particular evaluation of the number of coincidences in two-particle correlations and the dispersion of the corresponding probability distribution. These explicitly reflect fermion exchanges between components of composite bosons. We find that the number of coincidences in second-order correlations is enhanced compared to elementary bosons, so is the variance of the field. However, main statistical signatures are found to be preserved. Indeed, the extension of number-states to composite bosons still presents a strong quantum character, with a Mandel $Q$-parameter close to (-1).  

To highlight these effects in the simplest way, we have first considered $N=2$ composite bosons. From results obtained for just $N=2$ cobosons, we can use a rather intuitive "rule of the thumb" to determine the number of coincidences in two-particle correlations as well as the $Q$-parameter for arbitrary $N$. The corresponding expressions, guessed from the $N=2$ results, are thereafter confirmed by $N$-body calculations. These are done following a procedure we recently proposed \cite{report} to handle fermion exchanges between composite bosons, exactly. 

In general, our analysis confirms that the statistics of bosonic fields is modified by the underlying fermionic components of composite bosons. Precisely, fermion exchanges correlate composite boson states. This, in particular, makes the mean value of the number-operator smaller for composite bosons than for elementary ones. Therefore, coherence properties of composite bosons can not be directly deduced from their elementary boson counterparts. For that purpose, the derivation of operators peculiar to composite bosons, in order to possibly define "number-states" and "coherent-states", would be highly valuable.
\section{Appendix}

This appendix is dedicated to the calculation of $\langle B_0^\dag D_{00}B_0\rangle_N$ in a compact form. For that, we first use the commutator (\ref{eq5}). This allows us to write
\begin{equation}
\langle B_0^\dag D_{00}B_0 \rangle_N = \langle D_{00}B_0^\dag B_0 \rangle_N -2\langle L_2^\dag B_0\rangle_N\label{eqA1}.
\end{equation}
The part $\langle\psi_N|D_{00}$ in the first term is obtained by mixing Eqs.(\ref{eq16}) and (\ref{eq20}) . This leads to
\begin{equation}
D_{00}|\psi_N\rangle = 2NL_2^\dag|\psi_{N-1}\rangle= 2|\psi_N\rangle -\frac{2}{N+1}B_0|\psi_{N+1}\rangle.\label{eqA2}
\end{equation}
By using Eq. (\ref{eq22}), it is then easy to show that the first term of Eq.(\ref{eqA1}) also reads as
\begin{equation}
\langle D_{00}B_0^\dag B_0 \rangle_N = -2\Delta_N^{(1)}-2\frac{N(N-1)}{N+1}\Delta_N^{(2)}.\label{eqA3}
\end{equation}

In the second term of Eq.(\ref{eqA1}), we use Eq.(\ref{eq20}) for $B_0|\psi_N\rangle$ and then the same Eq.(\ref{eq20}), but for $L_2^\dag |\psi_{N-1}\rangle$. This leads to
\begin{equation}
\langle L_2^\dag B_0 \rangle_N = -\Delta_N^{(1)}-N(N-1)R,
\end{equation}
where $R$ is defined by
\begin{equation}
R = \frac{\langle \psi_N|L_2^{\dag 2}|\psi_{N-2}\rangle_N}{\langle \psi_N|\psi_N\rangle_N}. 
\end{equation}

All this allows us to rewrite Eq.(\ref{eqA1}) as 
\begin{equation}
\langle B_0^\dag D_{00}B_0 \rangle_N = -\frac{2N(N-1)}{N+1}\Delta_N^{(2)}+2N(N-1)R.\label{eqA4}
\end{equation}

To get $R$ in a compact form is considerably more difficult. Since it contains two Pauli scattering at least, through the two $L_2^\dag$ operators, we can already say that its leading term must be in $(a/L)^{2D}$. This is going to give a $\eta^2$ contribution to $g_N^{(2)}$, negligible compared to the dominant term of $g_N^{(2)}$, expected to be in $\eta$. Nevertheless, let us now show, for completeness, how a compact expression of $R$ can be obtained.

The trick is to note that, for $L_2^\dag$ given in Eq.(\ref{eq6}), we have, using Eq.(\ref{eq15})
\begin{eqnarray}
L_2|\psi_N\rangle = \sum_n \lambda(^{0 \hspace{.15cm} n}_{0\hspace{.15cm} 0})\left(N\delta_{n0}|\psi_{N-1}\rangle -N(N-1)\sum_m\lambda(^{m \hspace{.15cm} 0}_{n\hspace{.15cm} 0})B_m^\dag |\psi_{N-2}\rangle \right)\nonumber\\
= N\lambda_2|\psi_{N-1}\rangle -N(N-1)L_3^\dag|\psi_{N-2}\rangle,\label{eqA5}
\end{eqnarray}
 where $L_3^\dag$ is nothing but the operator defined in Eq.(\ref{eq7}) since $\sum\lambda(^{0 \hspace{.15cm} n}_{0\hspace{.15cm} 0})\lambda(^{m \hspace{.15cm} 0}_{n\hspace{.15cm} 0})$ is the three-body exchange scattering with three cobosons 0 on the right and two cobosons 0 plus one coboson {m} on the left, as readily seen from Figure 1.(b). By using the above equation for $\langle\psi_N|L_2^\dag$ but Eq.(\ref{eq20}) for $L_2^\dag|\psi_N\rangle$, it becomes easy to show that
\begin{equation}
R= \frac{\lambda_2}{N-1}\frac{F_{N-1}-F_N}{F_N}+\frac{\langle \psi_{N-1}|L_3|\psi_N\rangle-N\langle\psi_{N-2}|L_3|\psi_{N-1}\rangle}{\langle\psi_N|\psi_N\rangle}\label{eqA6}.
\end{equation}
 
The next step is to calculate $\langle\psi_{N-2}|L_3|\psi_{N-1}\rangle$. This is done by again using Eq.(\ref{eqA5}) for $\langle\psi_{N-2}|L_3$ and then Eq.(\ref{eq20})
for $L_2^{\dag}|\psi_{N-1}\rangle$. This leads to 
\begin{equation}
\langle\psi_{N-2}|L_3|\psi_{N-1}\rangle = \frac{\lambda_2}{N-1}\langle\psi_{N-1}|\psi_{N-1}\rangle -\frac{1}{N^2(N-1)}\langle\psi_N|\psi_N\rangle+\frac{1}{N^2(N^2-1)}\langle \psi_{N+1}|\psi_{N+1}\rangle,\label{eqA7}
\end{equation}
So that we ultimately find the expression of $R$ given in Eq.(\ref{eq30}). By using the values of $\Delta_N^{(1)}$ and  $\Delta_N^{(2)}$ in the small density limit, given above, we can show that $R\simeq \lambda_2^2$. The correction to $\langle B_0^\dag D_{00}B_0 \rangle_N$ induced by this $R$ term, is thus found to be of the order of $\eta^2$, as expected.

\end{document}